\documentclass[a4paper,fleqn]{cas-sc}

\usepackage[authoryear,longnamesfirst]{natbib}
\usepackage{amsmath}
\usepackage{autobreak}
\usepackage{amsfonts}
\usepackage{diagbox}
\usepackage{multirow}
\usepackage{makecell}
\usepackage{array}
\usepackage{booktabs}
\usepackage{bm}
\usepackage{graphicx}
\usepackage{algorithm}
\usepackage{subfigure}
\usepackage{caption}
\usepackage{algpseudocode}

\newcommand{\nop}[1]{}

\def\tsc#1{\csdef{#1}{\textsc{\lowercase{#1}}\xspace}}
\tsc{WGM}
\tsc{QE}

\begin{document}
\let\WriteBookmarks\relax
\def\floatpagepagefraction{1}
\def\textpagefraction{.001}

\shorttitle{}    

\shortauthors{}  

\title[mode = title]{A Time-varying Shockwave Speed Model for Trajectory Reconstruction using Lagrangian and Eulerian Observations}  



%

\author[1]{Yifan Zhang}[auid=000, orcid=0000-0001-8882-4327]
\cormark[1]


\ead{yifan.zhang@ivt.baug.ethz.ch}



\author[1]{Anastasios Kouvelas}[auid=001, orcid=0000-0003-4571-2530]
\ead{kouvelas@ethz.ch}

\author[1]{Michail A. Makridis}[auid=002, orcid=0000-0001-7462-4674]
\ead{michail.makridis@ivt.baug.ethz.ch}

\affiliation[a]{organization={Institute for Transport Planning and Systems, ETH Zurich},
            city={Zurich},
            postcode={8093}, 
            country={Switzerland}}

\cortext[1]{Corresponding author}



\begin{abstract}
Inference of detailed vehicle trajectories is crucial for applications such as traffic flow modeling, energy consumption estimation, and traffic flow optimization. Static sensors can provide only aggregated information, posing challenges in reconstructing individual vehicle trajectories. Shockwave theory is used to reproduce oscillations that occur between sensors. However, as the emerging of connected vehicles grows, probe data offers significant opportunities for more precise trajectory reconstruction. Existing methods rely on Eulerian observations (e.g., data from static sensors) and Lagrangian observations (e.g., data from probe vehicles) incorporating shockwave theory and car-following modeling. Despite these advancements, a prevalent issue lies in the static assignment of shockwave speed, which may not be able to reflect the traffic oscillations in a short time period caused by varying response times and vehicle dynamics. Moreover, energy consumption estimation is largely ignored. In response, this paper proposes a novel framework that integrates Eulerian and Lagrangian observations for trajectory reconstruction. The approach introduces a calibration algorithm for time-varying shockwave speed. The calibrated shockwave speed of the CV is then utilized for trajectory reconstruction of other non-connected vehicles based on shockwave theory. Additionaly, vehicle and driver dynamics are introduced to optimize the trajectory and estimate energy consumption. The proposed method is evaluated using real-world datasets, demonstrating superior performance in terms of trajectory accuracy, reproducing traffic oscillations, and estimating energy consumption. 
\end{abstract}

\begin{highlights}
\item Integrate Lagrangian and Eulerian observations to reconstruct trajectories.
\item Calibrate time-varying short-term shockwave speeds using the two types of data.
\item Reconstruct trajectories for non-connected vehicles based on shockwave theory.
\item Optimize trajectories by adding vehicle dynamics for better energy estimation.
\item Evaluation on real-world datasets shows excellent performances from several aspects.

\end{highlights}


\maketitle

\section{Introduction} \label{sec:intro}


Vehicle trajectory information with a high sampling frequency plays a key role in many domains, including but not limited to traffic flow modeling, energy consumption estimation, and traffic flow optimization~\citep{daganzo1997fundamentals,ZHANG2022103926,LI2020225,GUO2019313,FIORI2019275,he2020energy}. Collecting such granular information requires either a dense network of roadside sensors or transmission of location data from every individual vehicles. However, both approaches are financially and laboriously burdensome and maybe even infeasible in practice. In light of these constraints, it is crucial but challenging to infer the complete set of trajectories given limited sensing data rather than observing all of them directly. Thus, accurate reconstruction of trajectories with a minimal amount of Lagrangian observations provides a clear and significant contribution toward mixed traffic conditions.

Some methods are tailored to reconstruct trajectories using fixed sensors, including loop detectors, video cameras, radars, etc., which are deployed at a fixed location on the roadside to detect passing vehicles. This type of sensing data typically provides the information at an aggregated level, such as vehicle counts, average speed, and instantaneous velocities of passing vehicles, which is known as Eulerian observation. The data can be adopted to analyze and optimize the traffic flow from macroscopic views~\citep{948643,doi:10.3141/2260-14}, estimate travel time~\citep{COIFMAN2002351}, and reconstruct trajectory~\citep{COIFMAN2002351,fixed_sensor1,fixed_sensor2}. The shockwave theory is usually applied to this type of data to infer vehicles' trajectory, given that the traffic state along the shockwave is the same. Conventionally, the traffic state is assumed to propagate in the traffic flow at a fixed speed, referred to as the shockwave speed. Although the value of shockwave speed is constant on average over a long time period, the fact is that the traffic state may not propagate at the same speed in the short time period due to the existence of varying response times and vehicle dynamics~\citep{traffic_oscillation_freq_amp}, leading to sub-optimal reconstructed trajectories.


With the deployment of connected vehicles (CVs), mobile sensing data from CVs that report location information through vehicular networks, termed Lagrangian observation, largely contributes to trajectory reconstruction.  The collected trajectory data includes temporal and spatial data of these CVs, providing a rich information to analyze human driving behaviors~\citep{makridis2023characterising,SUAREZ2022103282,WANG2020135}, estimate fundamental diagram (FD)~\citep{SEO201940,LI2022103458}, and reconstruct the trajectory of vehicles~\citep{MONTANINO201582,https://doi.org/10.1049/itr2.12294,traj_smooth}. The trajectory of the CVs can be enhanced with a high accuracy using the collected data through interpolation curves~\citep{doi:10.1061/9780784479896.023}, particle filters~\citep{9027814}, maximum likelihood estimation~\citep{mobile_sensor1}, etc. The trajectory of non-connected vehicles (NCVs) is inferred by essentially replicating the trajectory of CVs based on Newells model, when the shockwave/backward speed in the model is a constant value~\citep{MONTANINO201582,https://doi.org/10.1049/itr2.12294}. However, such a manner may not be able to characterize realistic traffic scenarios, given the inherent variability in vehicle and driver dynamics, as previously discussed.

To overcome the above challenges of separately using Lagrangian or Eulerian observations, more attention is attracted to integrating both for reconstructing trajectories. Specifically, most of the methods~\citep{traj_recon_hybrid1,chen2022integrated,mehran2012implementing} are developed by deriving some reference points or candidate trajectories using one type of data and then optimizing them using the other type of data. Although the performance has been largely improved by the above hybrid methods, the following issues still need to be addressed. (1) Prevalent models often adhere to a fixed shockwave speed~\citep{chen2022integrated, mehran2012implementing}, while in reality, it has been observed to vary~\citep{futuretransp3040063,LEI2014453}. A constant value of shockwave speed implies that independently of the current traffic conditions or disturbances the propagation of oscillations follows a consistent pattern which is unrealistic~\citep{traffic_oscillation_freq_amp}. (2) Efforts to optimize time-varying shockwave speeds, as seen in~\citep{traj_recon_hybrid1}, encounter limitations when there are fewer than two vehicles affected by the same shockwave. 
(3) Reconstructed trajectories can capture traffic flow dynamics but they are still unreliable for energy consumption estimations where a realistic distribution of observed accelerations is essential~\citep{10159555}. (4) Calibration of car-following models for simulation will not be feasible when instantaneous vehicle dynamics are not realistic~\citep{8489919,HE2022103692,ZHENG2023104151}.

To this end, we design a new hybrid method based on the trajectories of a limited number of CVs and the instantaneous speed of vehicles collected by a loop detector. Specifically, to address the issues associated with fixed shockwave speed, we propose an innovative algorithm to calibrate the time-varying shockwave speed using Lagrangian and Eulerian observations. Given the calibrated shockwave speed, we develop an algorithm to reconstruct the reference points for NCVs. Moreover, to fill the gap in energy consumption estimation, we apply a vehicle movement model, that accounts for vehicle and driver dynamics, to optimize the reference points and generate the final trajectory of NCVs. In this paper, we mainly focus on highway scenarios without considering the disturbance of traffic signals and intersections. Our main contributions are summarized as follows.
\begin{itemize}
    \item We propose a new algorithm to calibrate the time-varying shockwave speed using the fixed sensor data and mobile sensor data. Specifically, our calibration algorithm is designed based on shockwave theory-compliant trajectory reconstruction methods. The algorithm aims to determine time-varying shockwave speed values that minimize the error between the reconstructed trajectory and the ground truth. We design and implement a Monte Carlo sampling-based method to achieve the aforementioned goal.
    \item We then design a trajectory reconstruction method using the time-varying shockwave speed and the data collected by fixed sensors. Each CV is considered as a leading vehicle, and our objective is to reconstruct the trajectory of the following vehicle of the CV. We derive reference points for the following vehicle using the CV's calibrated results according to shockwave theory.  These reference points are further optimized to reconstruct the trajectory, which is achieved by applying the microsimulation free-flow acceleration model (MFC) to fulfill the constraint of vehicle and driver dynamics.
    \item We evaluate and validate our method using real-world datasets. We compare with several baseline models in terms of time headway accuracy, speed accuracy, fuel consumption accuracy, CO$_2$ emission accuracy, and the capability of reproducing traffic oscillations. The experiment results show that our method achieves better performance compared with other baseline models, especially in the case of a low CV penetration rate.
    \item We further introduce the concept of a reconstructed spatial-temporal area to assess the contribution of CVs with different average speeds. Our findings reveal that a lower average speed can contribute more to reconstructing a larger spatial-temporal area. Moreover, we explore the relationship between the amount of required information for calibration and the average speed of the CV. The results show that the average speed does not affect the amount of required information.
\end{itemize}

The paper is organized as follows. In Section~\ref{sec:literature_review}, we review the literature on trajectory reconstruction and shockwave speed estimation. We first clarify the background and formulate the problem in Section~\ref{sec:background}. We then present the basic shockwave theory-based method and our method for calibrating time-varying shockwave speed in Section~\ref{sec:calibration}. The reconstruction process, including deriving reference points and reconstructing trajectory, is elaborated on in Section~\ref{sec:recon}. We then show the experiment results of our method compared with baseline models in Section~\ref{sec:experiment}. In Section~\ref{sec:discussion}, we further analyze the contribution of CVs in reconstructing trajectories and the amount of required information for calibrating time-varying shockwave speed. We finally conclude our paper and discuss the future work in Section~\ref{sec:conclusion}.

\section{Literature Review} \label{sec:literature_review}
Trajectory reconstruction has been studied for decades using different types of data and methods. The data from fixed sensors is largely used to reconstruct trajectories based on FIFO and shockwave theory. In recent years, the trajectory of some connected vehicles has also been widely applied to improve the reconstruction accuracy, which is regarded as the data from mobile sensors. In the rest of the section, we mainly review the literature about different trajectory reconstruction methods. Moreover, as most of the methods including ours are based on the shockwave theory, where the shockwave speed is one of the most important parameters, we also review the literature on shockwave speed estimation.

\subsection{Trajectory Reconstruction}
The method of trajectory reconstruction is mainly differentiated based on the type of data it leverages. 

Fixed sensors like loop detectors are very common in modern cities but they provide flow and potentially speed measurements at specific locations. Given the arrival time and velocity of vehicles collected by a fixed loop detector, \citep{COIFMAN2002351} firstly proposed to reconstruct the trajectory using the shockwave theory. Specifically, the vehicle's trajectory is reconstructed by concatenating the collected velocity of the following vehicles under the assumption that the velocity dissipates along the road at a shockwave speed. \citep{fixed_sensor1,fixed_sensor2} proposed to use data from multiple loop detectors. \citep{fixed_sensor1} Compared with~\citep{COIFMAN2002351}, \citep{fixed_sensor1} developed a method to calculate the linear velocity instead of applying the collected velocities as constant values to reconstruct the trajectories. \citep{fixed_sensor2} proposed filtered inverse speed-based (FISB) to construct a speed map for the reconstructed spatial-temporal space. The free flow and congestion are separately considered to estimate the speed of individual vehicles and further reconstruct trajectories. 


The second type is the data from mobile sensors. Usually, this type of data is reported by the individual CVs. One of the objectives of using this type of data is to reconstruct the trajectory of CVs since the frequency of the reported data from CVs may be too low to get enough information. Thus, many probabilistic methods are proposed to improve the resolution of the trajectory~\citep{mobile_sensor1,mobile_sensor2,9027814}. Moreover, trajectory reconstruction of NCVs is also one of the objectives. Usually, the historical trajectory of NCVs is reconstructed by replicating the trajectory of CVs using Newell-based models~\citep{MONTANINO201582,https://doi.org/10.1049/itr2.12294}. Besides, a huge amount of effort is made to reconstruct the future trajectory of the vehicle with the machine learning techniques~\citep{traj_pred}. In the rest of this paper, we mainly discuss reconstructing the fully sampled historical trajectory of NCVs.

The third one is to integrate the data from both fixed sensors and mobile sensors. More and more researchers focus on leveraging these two types of data to improve overall accuracy. Specifically, \citep{mehran2012implementing} proposed a data fusion framework to reconstruct the trajectories at signalized intersections based on the kinematic wave theory and variational solution of~\citep{DAGANZO2005187,DAGANZO2005934}. The temporal-spatial space of the trajectory is represented as a cumulative surface in a three-dimensional space. The data collected by the loop detector is the height of the cumulative surface at the specific temporal-spatial point. The trajectory of the CV is interpreted as a contour on the surface of cumulative curves. The other trajectories can be reproduced based on the estimation of this surface. Based on~\citep{mehran2012implementing}, \citep{9716064} designed a new framework, which classifies the temporal-spatial space into four regions and applies a Particle Filter-based fusion method to estimate the trajectory of vehicles at signalized intersections. \citep{REY201979,traj_recon_hybrid1} proposed to estimate the trajectory of vehicles on multi-lane roads based on the data from multiple loop detectors and CVs. A new variable of a vehicle's order is defined to represent its cumulative flows at different times and locations. \citep{chen2022integrated} designed a new framework to integrate both micro and macro models. The macro models based on shockwave theory aim to generate the velocity contour map, which includes the reference velocity. The micro model produces several candidate trajectories according to the trajectory of CVs. The final trajectories are optimized using the velocity contour map and candidate trajectories. All of the above methods are implemented based on wave theory, where the shockwave speed is one of the most important parameters. 

The main issue of the above methods is that the value of the shockwave speed is usually set as a constant value, which implies homogeneous vehicle and driver dynamics in the traffic flow. However, it is unrealistic according to the real-world observations~\citep{traffic_oscillation_freq_amp}. Moreover, the reconstructed trajectory is derived through a mathematical formulation or connecting several reference points directly, which may not satisfy the vehicle dynamics and can not be used for estimating energy consumption.


\subsection{Shockwave Speed Estimation}
As mentioned above, shockwave speed estimation plays a key role in reconstructing individual vehicle trajectories. Shockwave speed is an important parameter in describing the traffic flow. Many researchers have proposed different methods to estimate the value of shockwave speed in actual traffic. \citep{SEO201940} performs a linear regression between the flow and density for each pair of probe vehicles to estimate the shockwave speed. The final shockwave speed is the mean of all the regression results. \citep{mehran2012implementing} estimates the shockwave speed using the maximum flow, jam density, and free flow speed, which is largely determined by the road infrastructure. \citep{ANUAR2017183,traj_recon_hybrid1} locate the inflection point in the congested state and calculate the shockwave speed for each shockwave. Although~\citep{traj_recon_hybrid1} estimate different shockwave speeds for different periods to reconstruct the trajectories, it requires the trajectory of at least two CVs in the same shockwave to estimate the shockwave speed, which largely limits the performance with a lower penetration rate of CVs.

In summary, the trajectory reconstruction based on shockwave theory using fixed sensor and mobile sensor data has been widely studied. However, the challenges, including estimating and applying dynamic shockwave speeds, still remain unsolved. Moreover, the evaluation of the reconstructed trajectory usually focuses on the accuracy of the trajectory and largely ignores the energy aspects.

\section{Background and Problem Formulation} \label{sec:background}
\begin{figure}
    \centering
    \includegraphics[width=0.75\textwidth]{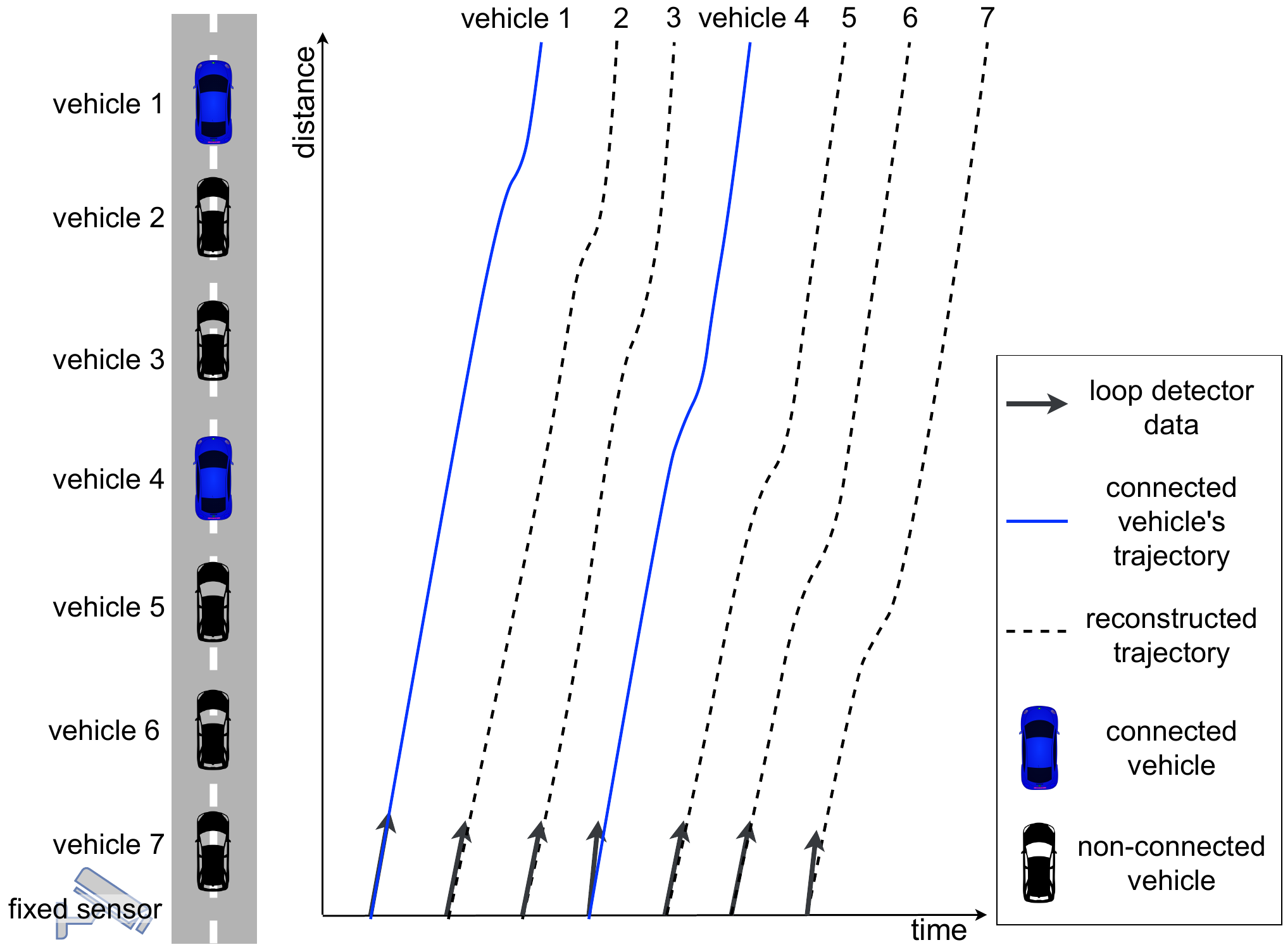}
    \caption{Illustration of the loop detector's information and connected vehicles' information.}
    \label{fig:loop_det_CV}
\end{figure}

We mainly consider the two following types of data to reconstruct the trajectories. The first one contains the arrival time and instantaneous speed of vehicles, collected by a fixed sensor deployed at the start of a road segment. The second one includes the trajectory of CVs, which are distributed randomly in the traffic flow at a lower penetration rate. As shown in Fig.~\ref{fig:loop_det_CV}, the fixed sensor is deployed at the start point of the road, and the blue vehicles are CVs. The objective is to use the above data to reconstruct the trajectories of other vehicles, namely, the black ones in Fig.~\ref{fig:loop_det_CV}.

Denote $t_i$ and $v_i$ as the arrival time and instantaneous speed of $i^{th}$ vehicle, respectively, which is collected by the fixed sensor at position $x_0$. Denote $\Pi_i= \{x_i^\tau, \tau \in [t_i, T]\}$ as the trajectory of $i^{th}$ vehicle, where $x_i^\tau$ is the longitudinal position of $i^{th}$ vehicle at time $\tau$ and $T$ is the end time of the trajectory. Denote $I$ as the union of all vehicles and $J$ as the union of all CVs, where $J \subset I$ and $|J| < |I|$. The problem of reconstructing the unknown trajectories is formulated as Eq.~\ref{eq:prob}.
\begin{equation} \label{eq:prob}
    \Pi_i = f(\Pi_J, t_I, v_I),\ \mathrm{where}\ i \in I \ \& \ i \notin J .
\end{equation}
We aim to derive the function $f(\cdot)$ to reconstruct $\{\Pi_i, i \in I \ \& \ i \notin J\}$ given $\Pi_J, t_I, $ and $v_I$.

\section{Time-varying Shockwave Speed} \label{sec:calibration}
The shockwave theory is the basis for reconstructing trajectories using different sources of data. Current methods usually assume a constant value for the parameter shockwave speed, which is not capable of handling dynamic and real traffic scenarios~\citep{futuretransp3040063,LEI2014453}. To address this issue, we propose to calibrate the time-varying shockwave speed for trajectory reconstruction using the data from a double loop detector and CVs. In the rest of the section, we first introduce the preliminaries of shockwave theory and trajectory reconstruction. We then present our algorithm to calibrate the time-varying shockwave speed values on the basis of the shockwave theory.

\subsection{Basics of Shockwave Theory}
The traffic shockwave is caused by some disturbances and propagates backward to the following vehicles, which is firstly proposed by Lighthill, Whitham and Richards (LWR)~\citep{lighthill1955kinematic}.
LWR model describes the relation between flow and density, which vary with location and time as shown in Eq.~\ref{eq:LWR}.
\begin{equation} \label{eq:LWR}
    \frac{\partial \mathrm{k}(x, t)}{\partial t} + \frac{\partial \mathrm{q}(x, t)}{\partial x} = 0,
\end{equation}
where $\mathrm{k}(x, t)$ is the density and $\mathrm{q}(x, t)$ is the flow at location $x$ and time $t$. Eq.~\ref{eq:LWR} can be further rewritten as Eq.~\ref{eq:shockwave_LWR}.
\begin{equation} \label{eq:shockwave_LWR}
    \frac{1}{w} \frac{\partial \mathrm{q}(x, t)}{\partial t} + \frac{\partial \mathrm{q}(x, t)}{\partial x} = 0,
\end{equation}
where $w = \frac{\partial \mathrm{q}}{\partial \mathrm{k}}$ is the shockwave speed. 

\begin{figure}
    \centering
    \subfigure[Newell fundamental diagram.]{
        \includegraphics[width=0.26\textwidth]{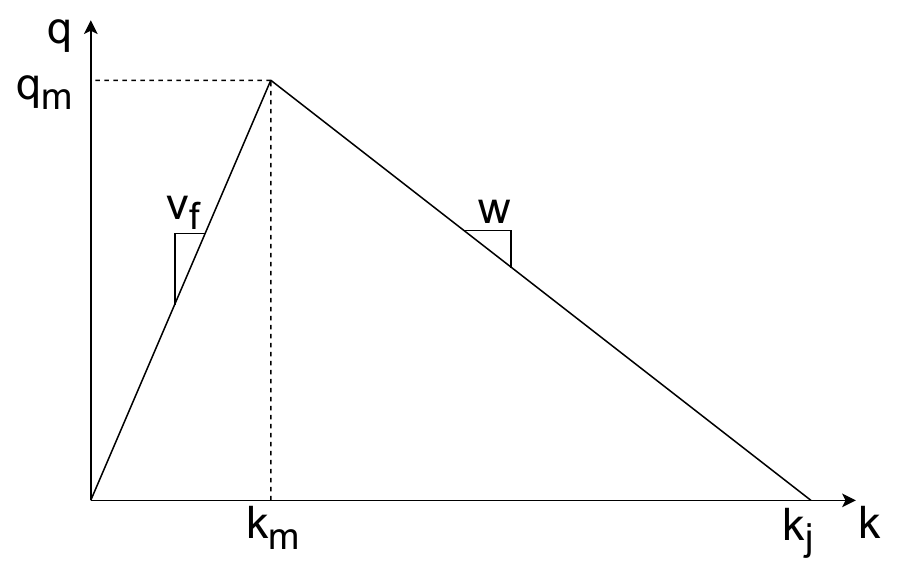}
        \label{fig:newell_fd}
    }
    \subfigure[An example of trajectory reconstruction based on shockwave theory.]{
        \includegraphics[width=0.70\textwidth]{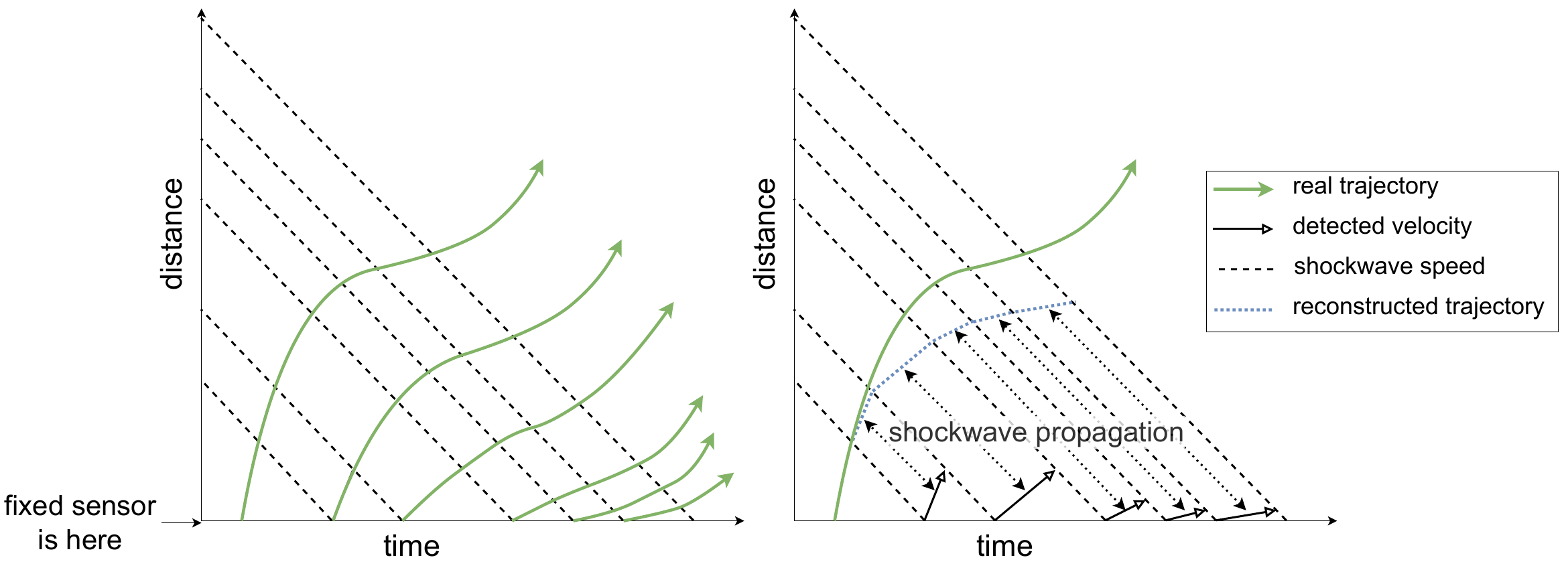}
        \label{fig:coifman_recon}
    }
    \caption{Trajectory reconstruction based on shockwave theory.}
\end{figure}

Following by LWR, Newell~\citep{NEWELL1993281} proposed a triangular fundamental diagram (FD) as Eq.~\ref{eq:newell}.
\begin{equation} \label{eq:newell}
    \mathrm{q}(\mathrm{k}) = \left\{
    \begin{aligned}
        & v_f \mathrm{k}, & \mathrm{k} \leq \mathrm{k}_m\\
        & w(\mathrm{k}_j - \mathrm{k}), & \mathrm{k} > \mathrm{k}_m
    \end{aligned}
    \right.
\end{equation}
where $v_f$ is the free flow speed, $\mathrm{k}_m$ is the road capacity, and $\mathrm{k}_j$ is the jam density. The triangle FD is shown as Fig.~\ref{fig:newell_fd}. The traffic states propagate among vehicles at the same speed on the same leg of the triangle, namely, $v_f$ for the free flow and $w$ for the congested flow. According to Eq.~\ref{eq:newell} and Fig.~\ref{fig:newell_fd},  the traffic states are the same in one shockwave, which makes traffic states predictable given the partial traffic state in the shockwave. The most important parameter to derive the traffic states is the shockwave speed, indicating how fast the traffic state of the leading vehicle is passed to the following ones.

\subsection{Trajectory Reconstruction Based on Shockwave Theory} \label{sec:coifman_traj_recon}
According to~\citep{COIFMAN2002351}, the trajectory of vehicles can be reconstructed given the arrival time and instantaneous speed of their following vehicles detected by the loop detector. Based on the above assumptions that the traffic state along the shockwave is the same, the trajectory of a vehicle is reconstructed by concatenating the trajectory segment of its following vehicles, as shown in Fig.~\ref{fig:coifman_recon}. Specifically, the trajectory segment of the following vehicle can be derived using the detected instantaneous speed of the vehicle. Then, these trajectory segments can be further concatenated following the shockwave propagation direction to construct the trajectory of the leading vehicle. 

\begin{figure}[!h]
    \centering
    \includegraphics[width=0.6\textwidth]{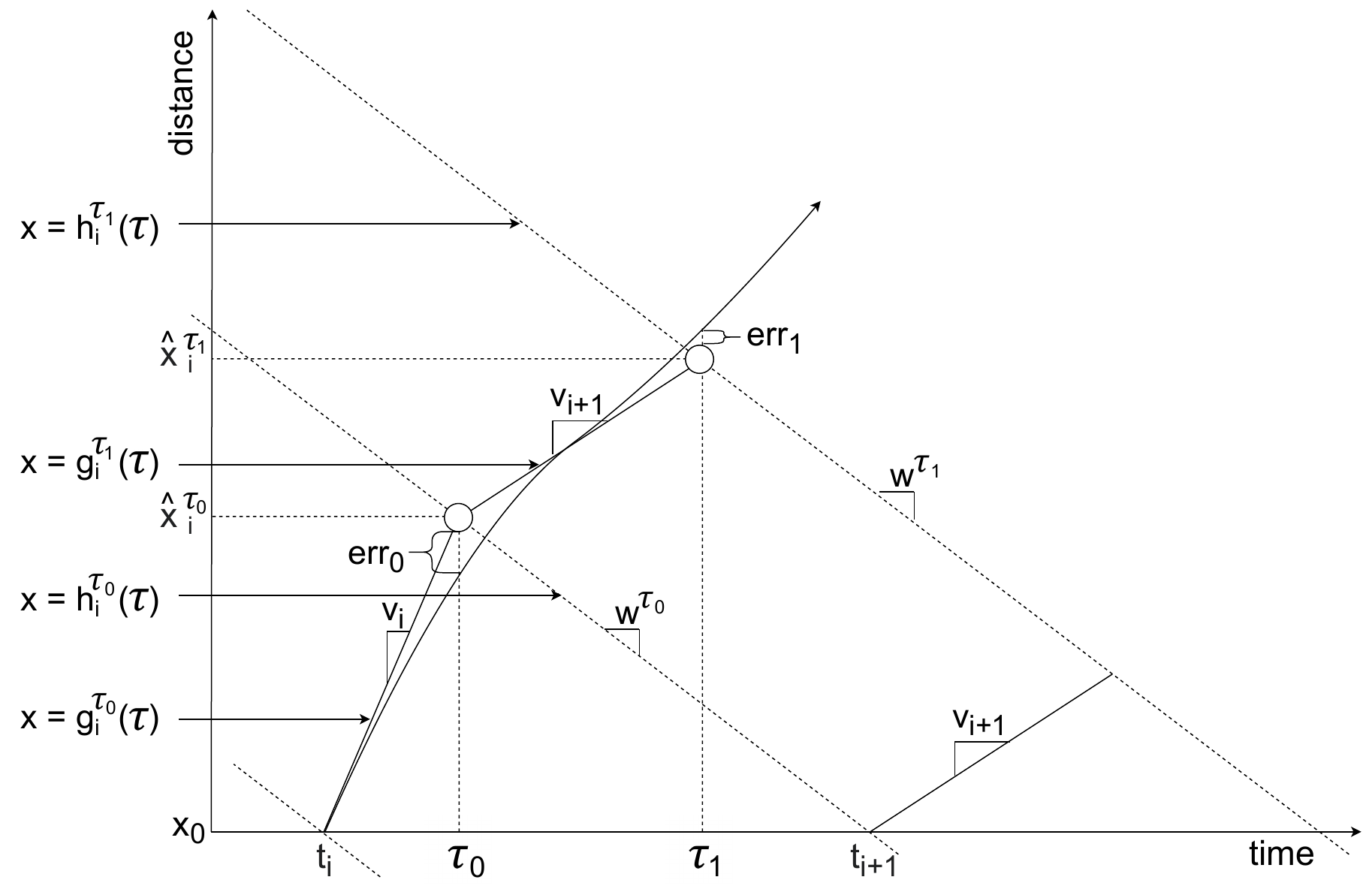}
    \caption{Process of shockwave speed calibration.}
    \label{fig:calibration}
\end{figure}

To simplify the reconstruction process, the reconstructed trajectory of each vehicle contains multiple line segments and each one is represented as a linear function in the temporal-spatial space, denoted as $g_i^{\tau_k}(\tau)$, where $i$ is the number of the vehicle and $k$ is the $k^{th}$ following vehicle of vehicle $i$. As shown in Eq.~\ref{eq:traj_line_func}, the slope of $g_i^{\tau_k}(\tau)$ is $v_{i+k}$, namely, the speed of $k^{th}$ following vehicle given the subject leading vehicle $i$, and $\hat{x}_i^\tau$ is the reconstructed trajectory point of vehicle $i$ at time $\tau$. Moreover, the shockwave propagation line that represents the propagation of speed $v_{i+k}$ is also modeled as a linear function $h_i^{\tau_k}(\tau)$ with slope $-w$, as shown in Eq.~\ref{eq:shockwave_line_func}. The notations are illustrated in Fig.~\ref{fig:calibration} for easy understanding.

\begin{equation} \label{eq:traj_line_func}
    g_i^{\tau_k}(\tau) = \left\{
    \begin{aligned}
        & v_i \tau + (x_0 - v_i t_i), & k = 0 \\
        & v_{i + k} \tau + (\hat{x}_i^{\tau_{k-1}} - v_{i + k} \tau_{k-1}), & k > 0
    \end{aligned}
    \right.
\end{equation}

\begin{equation} \label{eq:shockwave_line_func}
    h_i^{\tau_k}(\tau) =  -w \tau + (x_0 + w t_{i+k+1}).
\end{equation}

As mentioned above, the arrival time and instantaneous speed of vehicles at a fixed location $x_0$ are known. For obtaining the first trajectory segment of vehicle $i$, i.e., $k=0$, we calculate $(\tau_0, \hat{x}_i^{\tau_0})$ which is the point of intersection of $g_i^{\tau_0}(\tau)$ and $h_i^{\tau_0}(\tau)$. Thus, the first trajectory segment of vehicle $i$ is the line from point $(t_i, x_0)$ to $(\tau_0, \hat{x}_i^{\tau_0})$. Given $(\tau_0, \hat{x}_i^{\tau_0})$, $v_{i+1}$, $w$, and $(t_{i+2}, x_0)$, we can obtain $g_i^{\tau_1}(\tau)$ and $h_i^{\tau_1}(\tau)$ so that the second trajectory segment is the line from $(\tau_0, \hat{x}_i^{\tau_0})$ to $(\tau_1, \hat{x}_i^{\tau_1})$. As such, $(\tau_k, \hat{x}_i^{\tau_k})$, $g_i^{\tau_k}(\tau)$, $h_i^{\tau_k}(\tau)$ can be further derived to complete the trajectory of vehicle $i$ by repeating the above procedures. In the rest of the paper, the process of calculating the $k^{th}$ trajectory segment is abbreviated as reconstruction step $k$.

\begin{algorithm}[!h]
\caption{Time-varying Shockwave Speed Calibration}
\label{al:calibration}
\begin{algorithmic}[1]
\Require{Arrival time $\{t_i, i \in I\}$ and speed $\{v_i, i \in I\}$ collected by loop detector, trajectory of CVs $\{\Pi_j, j \in J\}$, upper bound and lower bound of shockwave speed $[w_{ub}, w_{lb}]$, number of samples $N$, accept thresholds $\epsilon$, maximum iterations $M$} 
\Ensure{Time-varying shockwave speed of CVs $\{W_j, j \in J\}$, where $W_j = \{w_j^{\tau_k}, k \in [0, 1, 2, ..., k_{end}]\}$}
\For{$j \in \mathcal{J}$}
\State $D = \{(t_i, v_i), \forall t_i > t_j \}$
\State Sort $D$ in ascending order of $t_i$
\State $k = 0$
\For{$(t_i, v_i) \in D$ do}
\State $iter = 0$
\While{True}
\State Sample $N$ samples from uniform distribution $U(w_{ub}, w_{lb})$
\State Using the samples $\mathbf{w}=\{w_n, n \in [1, 2, ..., N]\}_j^{\tau_k}$ to derive $g_j^{\tau_k}(\tau)$ and $\tilde{h}_j^{\tau_k}(\tau, w_n)$
\State Calculate the point of intersection $(\tau_k, \hat{x}_j^{\tau_k})_{w_n} = g_j^{\tau_k}(\tau) \otimes \tilde{h}_j^{\tau_k}(\tau, w_n)$
\State Measure errors $\textbf{err} = \{||(\tau_k, \hat{x}_j^{\tau_k})_{w_n}, (\tau_k, x_j^{\tau_k})||, n \in [1, 2, ..., N] \}$
\State $w^* = \mathop{\arg\min}_w \textbf{err}$
\If{$\mathrm{min}(\textbf{err}) < \epsilon$}
\State $w_j^{\tau_k} = w^*$
\State \textbf{break}
\ElsIf{$(\hat{x}_j^{\tau_k})_{w^*} < x_j^{\tau_k}$}
\State Increase $v_i$
\ElsIf{$(\hat{x}_j^{\tau_k})_{w^*} > x_j^{\tau_k}$}
\State Decrease $v_i$
\EndIf
\State $iter = iter + 1$
\If{$iter > M$}
\State $w_j^{\tau_k} = w^*$
\State \textbf{break}
\EndIf
\EndWhile
\State $k = k + 1$
\EndFor
\EndFor
\end{algorithmic}
\end{algorithm}

\subsection{Calibration Algorithm}

Several studies in the literature discuss the estimation of shockwave speed~\citep{SEO201940,traj_recon_hybrid1}. Some of them have proved that the value should not be fixed~\citep{traj_recon_hybrid1,ANUAR2017183}. However, the shockwave speed in Coifman's method and its variants is set as a constant value, which may lead to sub-optimal results. This motivated this work to calibrate time-varying shockwave speed values.

Besides the loop detector data, the trajectory data reported by CVs is leveraged to achieve this goal. The objective of calibration is to find the optimal shockwave speed value $w_j^{\tau_k}$ for CV $j$ at reconstruction step $k$, which minimizes the error between the ground truth trajectory and the reconstructed one, as shown in Fig.~\ref{fig:calibration}. The objective function is formulated as Eq.~\ref{eq:obj_calibration}.
\begin{equation} \label{eq:obj_calibration}
    w_j^{\tau_k} = \mathop{\arg\min}_w ||g_j^{\tau_{k}}(\tau) \otimes \tilde{h}_j^{\tau_{k}}(\tau), (\tau_k, x_j^{\tau_k})||,
\end{equation}
where the operation $\otimes$ is to calculate the point of intersection, $||\cdot, \cdot||$ is the distance between two points, and $\tilde{h}_j^{\tau_{k}}(\tau)$ is the time-varying variant of Eq.~\ref{eq:shockwave_line_func}. Specifically, the time-varying shockwave line function is illustrated in Eq.~\ref{eq:tv_shockwave_line_func}.

\begin{equation} \label{eq:tv_shockwave_line_func}
    \tilde{h}_j^{\tau_k}(\tau) = -w_j^{\tau_k} \tau + (x_0 + w_j^{\tau_k} t_{j+k+1}).
\end{equation}

The optimization problem in Eq.~\ref{eq:obj_calibration} can not be solved explicitly since the ground truth trajectory can not be represented using the enclosed mathematical formulation so that the point $(\tau_k, x_j^{\tau_k})$ can not be determined before knowing $\tilde{h}_j^{\tau_{k}}(\tau)$ and $g_j^{\tau_{k}}(\tau)$. However, the slope $-w_j^{\tau_k}$ of $\tilde{h}_j^{\tau_{k}}(\tau)$ is the unknown parameter to be optimized. Namely, $(\tau_k, x_j^{\tau_k})$ is indeterminate leading that $w_j^{\tau_k}$ in Eq.~\ref{eq:obj_calibration} can not be solved explicitly. Thus, we propose to solve this problem through the Monte Carlo sampling-based method. Specifically, for each reconstruction step $k$, we generate $N$ samples of $w_j^{\tau_k}$, denoted as $\{w_n, n \in [1,2,...,N]\}_j^{\tau_k}$. For each sampled value $w_n$ of $w_j^{\tau_k}$, we then calculate the point of intersection between the trajectory line $g_j^{\tau_{k}}(\tau)$ and the shockwave line $\tilde{h}_j^{\tau_{k}}(\tau, w_n)$. Given the point of intersection, we can derive the error between the point of intersection and the corresponding point on the ground truth trajectory, denoted as $\{err_n, n\in[1,2,...,N]\}$ corresponding to $\{w_n, n \in [1,2,...,N]\}_j^{\tau_k}$, respectively. Note that considering the reconstructed trajectory may have a longer travel time than the ground truth, we calculate the time difference under the same position between the ground truth point and the point of intersection. The error in the temporal dimension is independent of speed, which is better than the one in the spatial dimension for optimization. Given $\{err_n, n\in[1,2,...,N]\}$, the minimal error is compared with a pre-defined threshold $\epsilon$. The corresponding $w_n$ is kept for the reconstruction step $k$ if the above minimal error is smaller than $\epsilon$.

\begin{figure}[!h]
    \centering
    \includegraphics[width=0.55\textwidth]{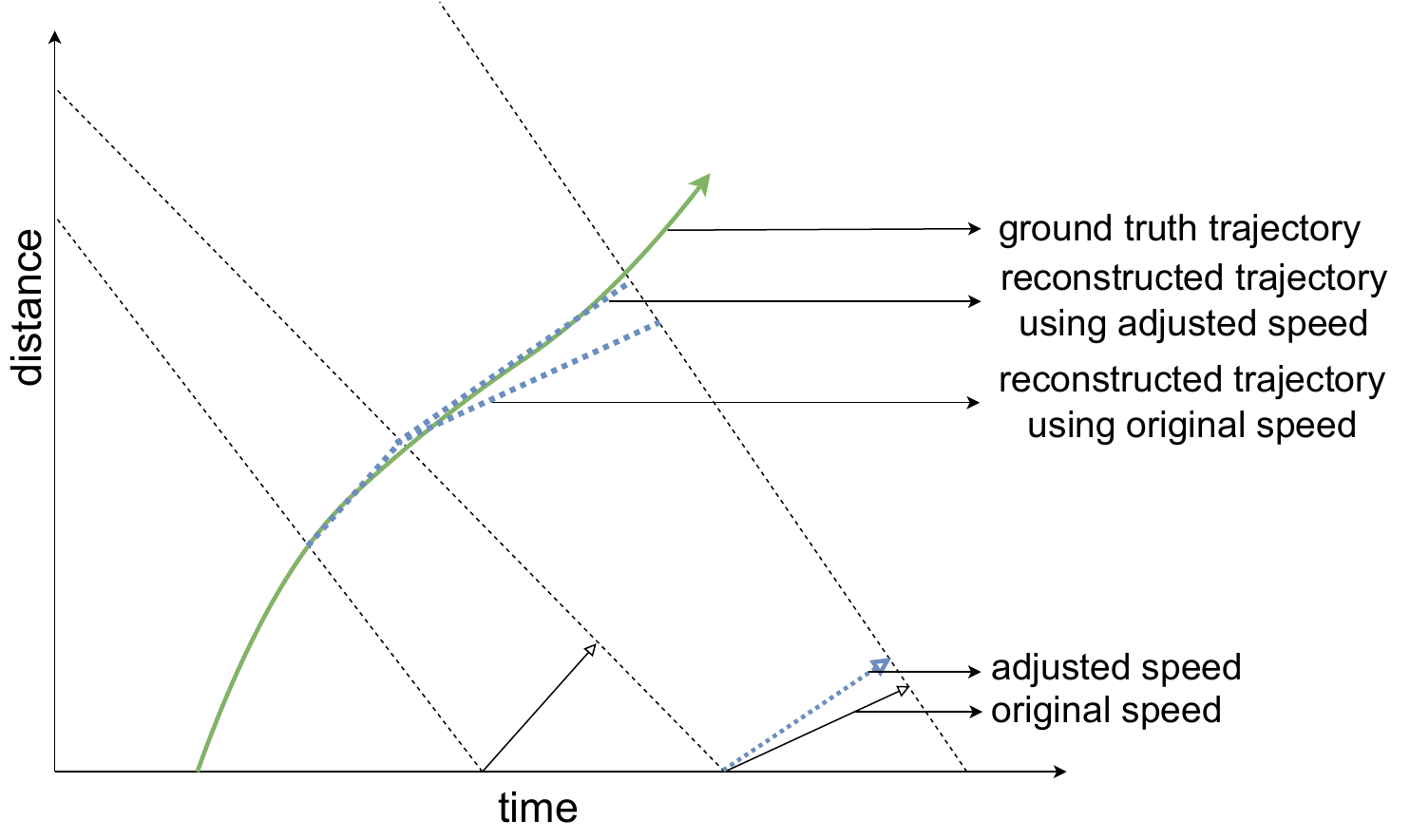}
    \caption{Adjust the speed collected by the loop detector.}
    \label{fig:ada_v_j}
\end{figure} 

However, as the value of $w_j^{\tau_k}$ is bounded by an upper and a lower bound, it could be possible that there is no such $w_n$ that can achieve such a small error. To solve this problem, we optimize the value of both $w_j^{\tau_k}$ and $v_{j+k}$. Translating the speed of the following vehicle to reconstruct the trajectory of the leading vehicle can be modeled as a quantification of the impact of the leading vehicle on the following ones. Nonetheless, the stochasticity of such impacts can not be ignored, leading to the aforementioned challenge of solving the optimal $w_j^{\tau_k}$. Thus, to well handle the stochasticity, we adjust the speed of the following vehicles to approximate the ground truth trajectory, as illustrated in Fig.~\ref{fig:ada_v_j}. Specifically, we compare the point of intersection $(\tau_k, \hat{x}_j^{\tau_k})$ with minimal error and the point $(\tau_k, x_j^{\tau_k})$ on the ground truth trajectory, when the minimal error can not satisfy the threshold $\epsilon$. The value of $v_{j+k}$ is increased/decreased if the reconstructed point is at the rear/front position of the ground truth point, as shown in Fig.~\ref{fig:ada_v_j}. The pseudo-code of the algorithm is illustrated in Algorithm~\ref{al:calibration}.

\section{Trajectory Reconstruction using Time-varying Shockwave Speed} \label{sec:recon}
Given the calibrated time-varying shockwave speed of the CV, we regard the CV as a leading vehicle to reconstruct the reference points for the NCVs which are the following vehicles of the CV. The reconstruction process is designed based on the method introduced in Section~\ref{sec:coifman_traj_recon}. However, simply connecting these reference points leads to a harsh trajectory, which can not fulfill the vehicle dynamics and thus can not be used to estimate energy consumption. Hence, we propose to design and apply a trajectory reconstruction method to filter the reference points and generate the final trajectory. In the rest of this section, we introduce the process of reconstructing the reference points and optimizing the reference points to reconstruct the final trajectory.

\subsection{Reference Points Reconstruction}
With the calibrated time-varying shockwave speed, it is simple to reconstruct the trajectory of CVs according to the process in Section~\ref{sec:coifman_traj_recon}. However, the trajectory reconstruction of other NCVs remains unsolved. To this end, we propose to reconstruct the reference points of these NCVs using the calibrated shockwave speed under the assumption that the shockwave speed does not change in a short time period. 

\begin{figure}[htb!]
    \centering
    \includegraphics[width=0.7\textwidth]{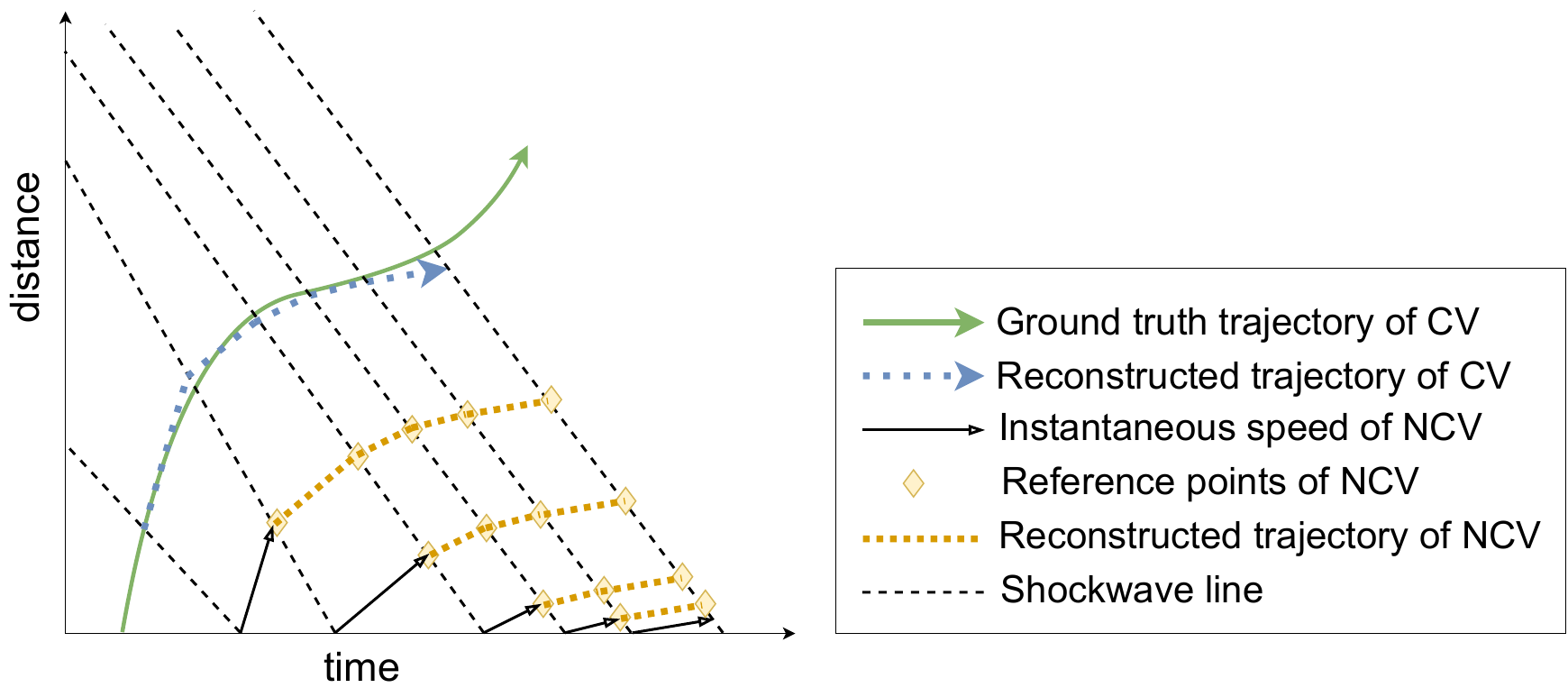}
    \caption{Trajectory reconstruction of other non-connected vehicles.}
    \label{fig:non_connected_traj_recon}
\end{figure}

Given the calibrated time-varying shockwave speed, which is denoted as $\{w_j^{\tau_k}, k \in [0,1,...,k_{end}], j \in J \}$, the representation of the shockwave line in Eq.~\ref{eq:shockwave_line_func} is rewritten as Eq.~\ref{eq:tv_shockwave_line_func}. The reference points of CVs can be reconstructed according to the procedures in Section~\ref{sec:coifman_traj_recon} using the calibrated shockwave speed, Eq.~\ref{eq:traj_line_func}, and Eq.~\ref{eq:tv_shockwave_line_func}. Considering that the traffic conditions do not change dramatically in a short time period, we use the calibrated shockwave speed of the closest leading CV for the NCVs to reconstruct their reference points. Specifically, for $j^{th}$ CV with calibrated time-varying shockwave speed, each shockwave speed value $w_j^{\tau_k}$ is paired with $v_{j+k}$ the speed of $k^{th}$ following vehicle, where the CV is considered as the leading vehicle. Thus, the reference points of $k^{th}$ following vehicle of the CV can be reconstructed through calculating the points of intersection $\{\tilde{h}_{j+k}^{\tau_m}(\tau, w_j^{\tau_{k+m}}) \otimes g_{j+k}^{\tau_m}(\tau), m \in [0, 1, ..., k_{end} - k] \}$, as shown in Fig.~\ref{fig:non_connected_traj_recon}. The stochasticity of the impact of the leading vehicle on the following ones through shockwave propagation is reflected by sampling $\tilde{v}_{j+k+m}$ from Gaussian distribution with mean $v_{j+k+m}$ and pre-defined variance $\sigma^2$ to derive $g_{j+k}^{\tau_m}(\tau)$. After calculating the point of intersection $(\hat{x}_{j+k}^{\tau_m'}, {\tau_m'})$, we need to validate it by comparing the position and time with the existing reference point $(\tau_{m'-1}, \hat{x}_{j+k}^{\tau_{m'-1}})$. This new reference point is added to reconstruct the trajectory if the position and time of this new point are greater than the existing point of the previous time step. The reconstruction process of reference points is presented in Algorithm~\ref{al:traj_recon}.

\begin{algorithm}
\caption{Reference Point Reconstruction for NCVs}
\label{al:traj_recon}
\begin{algorithmic}[1]
\Require{Arrival time $\{t_{j+k}, t_{j+k+1}, ..., t_{j+k_{end}}\}$ and speed $\{v_{j+k}, v_{j+k+1}, ..., v_{j+k_{end}}\}$ of the $(j+k)^{th}$ to $(j+k_{end})^{th}$ NCVs, the calibrated shockwave speed of $j^{th}$ CV $\{w_j^{\tau_0}, w_j^{\tau_1}, ..., w_j^{\tau_\kappa}, ..., w_j^{\tau_{k_{end}}}\}$, $j^{th}$ CV is the closest leading CV for its following $1^{st}$ to $\kappa^{th}$ NCVs, the location of loop detector $x_0$, standard deviation $\sigma$ } 
\Ensure{Reconstructed trajectory of $(j+k)^{th}$ NCV, where $k \leq \kappa$}
\State $\Pi_{j+k} = \{(t_{j+k}, x_0)\}$
\For{$m = k:\kappa$}
\State Sample $\tilde{v}_{j+m}$ from Gaussian distribution $N(v_{j+m}, \sigma^2)$
\State Derive $\tilde{h}^{\tau_{m-k}}_{j+k}(\tau)$ using Eq.~\ref{eq:tv_shockwave_line_func} and $w_j^{\tau_{m}}$
\State Derive $g^{\tau_{m-k}}_{j+k}(\tau)$ using Eq.~\ref{eq:traj_line_func} and $\tilde{v}_{j+m}$
\State Calculate the point of intersection $(\tau_{m-k}, \hat{x}_{j+k}^{\tau_{m-k}})=\tilde{h}^{\tau_{m-k}}_{j+k}(\tau) \otimes g^{\tau_{m-k}}_{j+k}(\tau)$
\If{$\tau_{m-k} > \tau_{m-k-1}$ \textbf{and} $\hat{x}_{j+k}^{\tau_{m-k}} > \hat{x}_{j+k}^{\tau_{m-k-1}}$}
\State Append $(\tau_{m-k}, \hat{x}_{j+k}^{\tau_{m-k}})$ in $\Pi_{j+k}$
\EndIf
\EndFor
\end{algorithmic}
\end{algorithm}

\subsection{Trajectory Reconstruction} \label{sec:traj_smooth}
Although the above reference points can be connected directly to produce the final trajectory of the vehicle, the trajectory is not realistic since it does not consider vehicle or driver dynamics. Moreover, the energy consumption estimated using such a trajectory is not reliable either. To this end, we propose a further step to reconstruct the trajectory using the reference points. The reconstructed trajectory is constrained by the vehicle and driver dynamics for a more realistic trajectory and better energy estimation. 

We apply a vehicle movement model to introduce vehicle and driver dynamics simultaneously. We choose the microsimulation free-flow acceleration model (MFC)~\citep{mfc_model} as the vehicle movement model. MFC is developed to capture the vehicle acceleration dynamics accurately and consistently. The driver dynamics are also considered in MFC by introducing the driver style parameter. MFC takes the desired speed as input to simulate the acceleration behaviors given the driver style and vehicle's physical parameters, such as the mass of the vehicle, the maximum tractive force of the vehicle, etc. As such, we consider the speed of the reference points as the desired speed and input to the MFC to obtain the acceleration behaviors. Specifically, the desired speed is first enhanced by the interpolation method to increase the frequency. The interpolated desired speed is then input to the MFC to derive acceleration behaviors, which are further used to reconstruct the final trajectory and speed.

\begin{figure}[!h]
    \centering
    \includegraphics[width=0.60\textwidth]{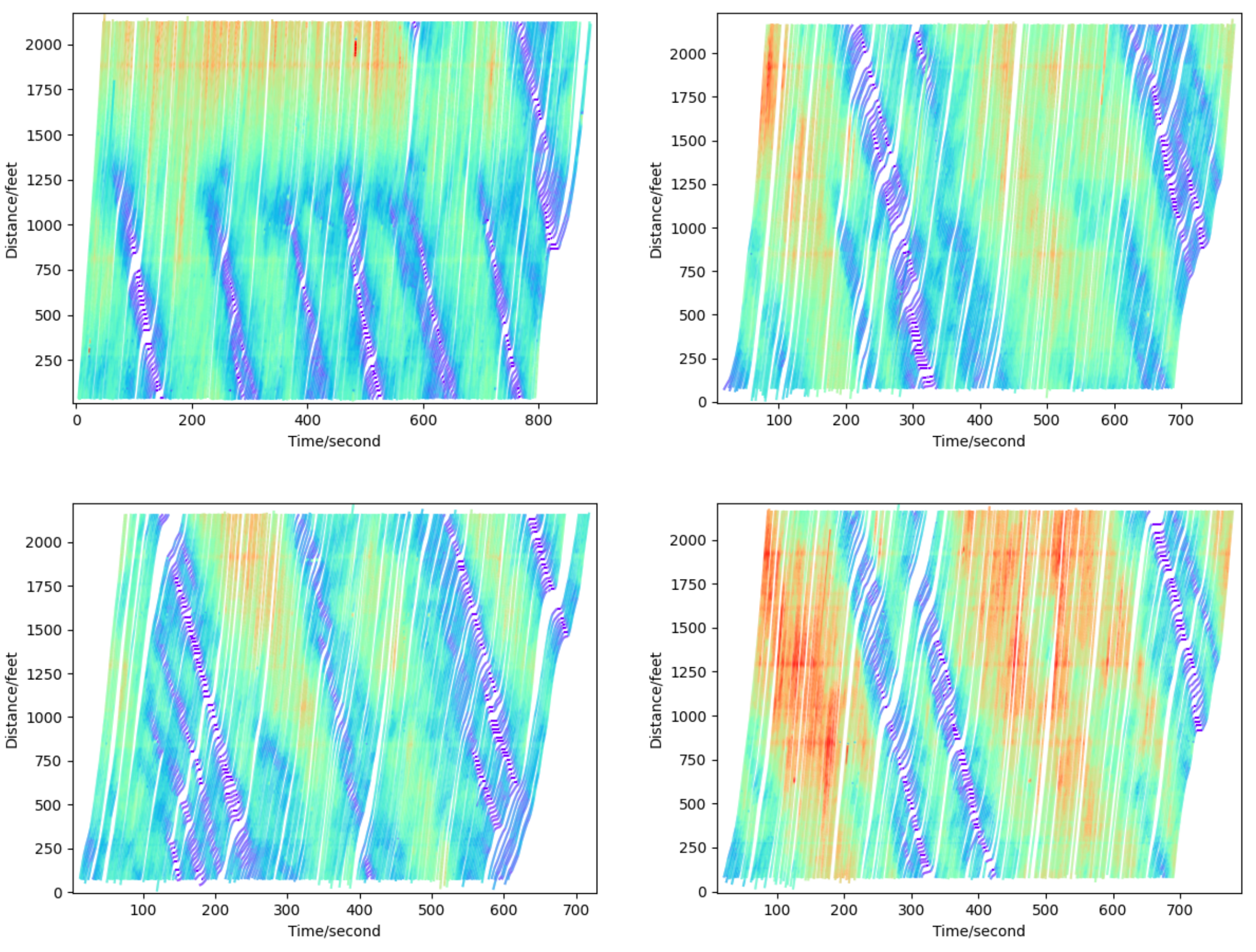}
    \caption{Examples of extracted trajectory data.}
    \label{fig:traj_examples}
\end{figure}

\begin{table}[!b]
    \centering
    \begin{tabular}{c|cc|cc|cc}
    \toprule
        \multirow{3}*{\diagbox{Model}{Metric}{Penetration rate}} & \multicolumn{6}{c}{$5\%$} \\
        \cline{2-7}
         & \multicolumn{2}{c|}{Time headway (s)}  & \multicolumn{2}{c|}{Speed (m/s)} & \multicolumn{2}{c}{Fuel consumption (L/100km)}  \\
         & avg. & std. & avg. & std. & avg. & std. \\
        \midrule
         Coifman & 5.61 & - & 3.86 & - & 0.37 & - \\
         Macro-micro & 3.82 & 0.27 & 2.86 & 0.19 & 0.32 & 0.06 \\
         Ours with polynomial & 2.65 & 0.17 & 2.74 & 0.18 & 0.65 & 0.06 \\
         Ours with MFC & \textbf{2.64} & 0.19 & \textbf{2.65} & 0.17 & \textbf{0.23} & 0.02 \\
        \hline
        \hline
        \multirow{3}*{\diagbox{Model}{Metric}{Penetration rate}} & \multicolumn{6}{c}{$10\%$} \\
        \cline{2-7}
         & \multicolumn{2}{c|}{Time headway (s)}  & \multicolumn{2}{c|}{Speed (m/s)} & \multicolumn{2}{c}{Fuel consumption (L/100km)}  \\
         & avg. & std. & avg. & std. & avg. & std. \\
        \midrule
         Coifman & 5.61 & - & 3.86 & - & 0.37 & -  \\
         Macro-micro & 2.27 & 0.13 & 2.70 & 0.06 & 0.28 & 0.03 \\
         Ours with polynomial & 2.24 & 0.11 & 2.74 & 0.11 & 0.66 & 0.03 \\
         Ours with MFC & \textbf{2.19} & 0.11 & \textbf{2.61} & 0.10 & \textbf{0.21} & 0.01 \\
        \hline
        \hline
        \multirow{3}*{\diagbox{Model}{Metric}{Penetration rate}} & \multicolumn{6}{c}{$15\%$} \\
        \cline{2-7}
         & \multicolumn{2}{c|}{Time headway (s)}  & \multicolumn{2}{c|}{Speed (m/s)} & \multicolumn{2}{c}{Fuel consumption (L/100km)}  \\
         & avg. & std. & avg. & std. & avg. & std. \\
        \midrule
         Coifman & 5.61 & - & 3.86 & - & 0.37 & -  \\
         Macro-micro & \textbf{1.72} & 0.12 & \textbf{2.15} & 0.07 & 0.24 & 0.02 \\
         Ours with polynomial & 2.07 & 0.08 & 2.71 & 0.06 & 0.63 & 0.02 \\
         Ours with MFC & 2.01 & 0.10 & 2.59 & 0.07 & \textbf{0.21} & 0.01 \\
        \bottomrule
    \end{tabular}
    \caption{Comparison of MAE of time headway, fuel consumption, and CO$_2$ among our model with different trajectory reconstruction methods and baseline models. }
    \label{tab:acc_comp}
\end{table}

\begin{figure}[!h]
    \centering
    \subfigure[Time headway MAE.]{
        \includegraphics[width=0.46\textwidth]{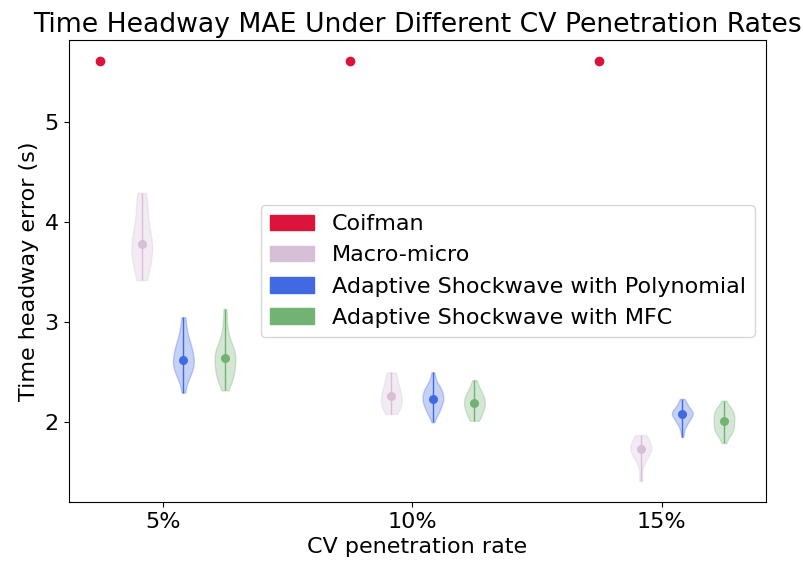}
        \label{fig:spacing_err}}
    \subfigure[Speed MAE.]{
        \includegraphics[width=0.47\textwidth]{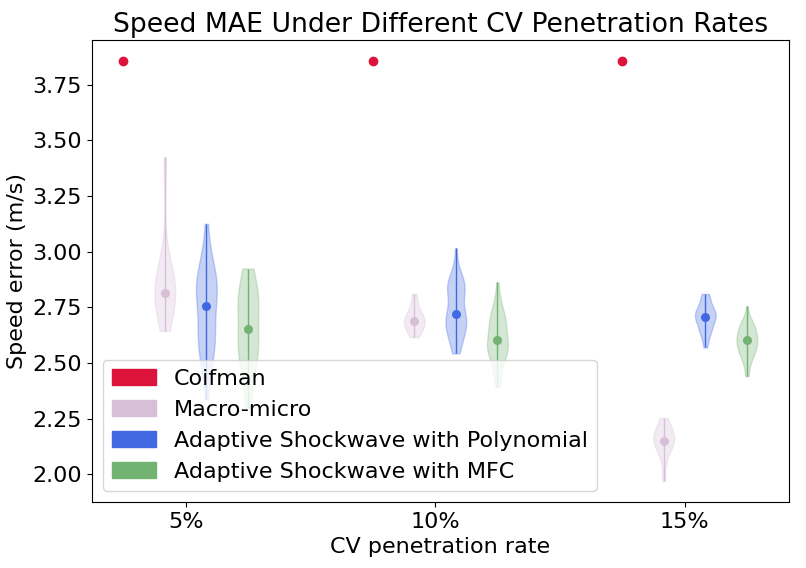}
    }
    \subfigure[Fuel consumption MAE.]{
        \includegraphics[width=0.48\textwidth]{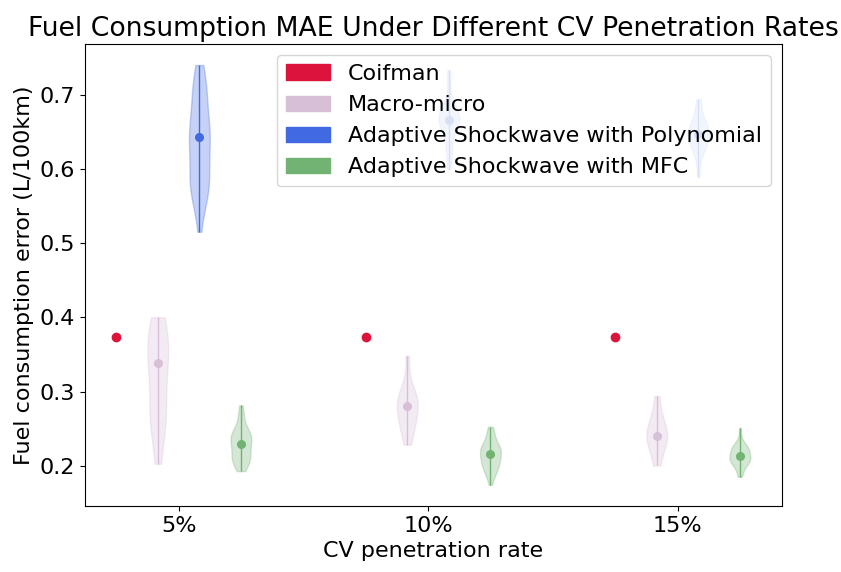}
        \label{fig:fc_err}}
    \subfigure[CO2 emission MAE.]{
        \includegraphics[width=0.47\textwidth]{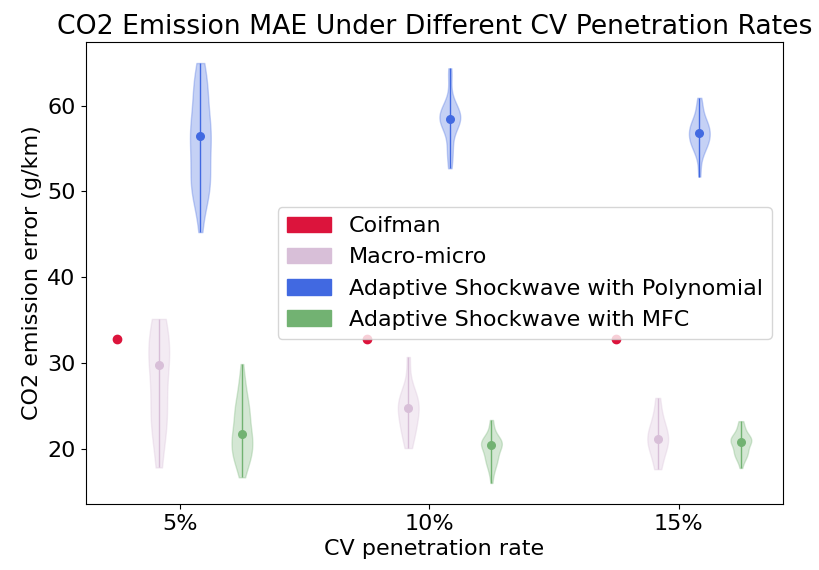}
        \label{fig:co2_err}}
    \caption{Accuracy comparison of time headway, fuel consumption, and CO$_2$ emission among baseline models and ours.}
    \label{fig:acc_comp}
\end{figure}

\section{Experiments} \label{sec:experiment}
To evaluate the performance of our proposed trajectory reconstruction method, we conduct the following experiments: (1) comparison with several SOTA baseline models in terms of trajectory reconstruction accuracy and energy consumption accuracy; (2) evaluation of the capability of capturing the traffic oscillation; (3) exploration the relationship between the accuracy of reconstructed trajectory and the penetration rate of CVs
The details are elaborated in the rest of this section.

\subsection{Data Preprocessing}
We use the HighD~\citep{highDdataset} and NGSIM~\citep{us101,i80} datasets to evaluate our method, which are collected on the highways of Germany and the US, respectively. They include the trajectory of vehicles of several minutes and in multiple lanes. We first separate each dataset into several sub-datasets according to lanes. Namely, the trajectory of vehicles on the same lane is formed as a new sub-dataset. Moreover, since our method is based on shockwave speed theory, we only use the sub-dataset that exists shockwave to evaluate the performance of our method. Only the car-following vehicles are kept without considering lane-change ones. The trajectories of some of the selected sub-dataset are shown in Fig.~\ref{fig:traj_examples}. We assume that a loop detector is deployed at the start position of the road segment, as shown in Fig.~\ref{fig:loop_det_CV}.  We collect the velocity of all of the vehicles that pass the loop detector. Some of the vehicles are defined as CVs with known trajectories, which are the blue ones in Fig.~\ref{fig:loop_det_CV}.



\subsection{Accuracy Comparison}

To assess the performance of our method, we choose the following baseline models: (1) Coifman's method~\citep{COIFMAN2002351} ; (2) macro–micro model~\citep{chen2022integrated}. We follow the values of parameters as mentioned in~\citep{chen2022integrated} to reproduce the reconstruction results. Additionally, we also conduct an ablation study on the selection of trajectory reconstruction method. Besides the aforementioned trajectory reconstruction in~\ref{sec:traj_smooth}, we apply a polynomial reconstruction method for comparison. Specifically, the reference points are reconstructed using a 5-degree polynomial optimized by least-square regression. We use mean absolute error (MAE) on time headway and speed to measure the accuracy of the reconstructed trajectory. To estimate the energy consumption and emission, we apply the simplified fuel consumption model~\citep{mfc_model} and CO$_2$ emission model, which is publicly available~\footnote{https://pypi.org/project/co2mpas-driver/} and is based on CO$_2$MPAS~\citep{co2mpas} model, developed by the European Commission. We compare the total amount of energy consumed on the same travel length between our model and the ground truth trajectory.

Moreover, we set different penetration rates, $5\%$, $10\%$, and $15\%$, to explore its impact on the accuracy. To achieve $5\%$ and $10\%$ penetration rates, we randomly select one vehicle as the CV with a known trajectory every twenty or ten vehicles, respectively. For $15\%$ penetration rate, we randomly select three vehicles as the CVs with known trajectory every twenty vehicles. For different penetration rates, we run 50 times and the CVs are selected randomly every time. As we have no information about the vehicle dynamics from the dataset, we randomly select an average vehicle model of European Class C~\citep{vehicle_class} to estimate fuel consumption and CO$_2$ emissions.

The comparison results of two baseline models and our model with different reconstruction methods are shown in Tab.~\ref{tab:acc_comp} and violin plot as Fig.~\ref{fig:acc_comp}. \textbf{\textit{In terms of time headway accuracy,}} ours with both polynomial and MFC reconstruction methods perform much better than the baseline models when the penetration rate is low, i.e., 5\% in our experiment. When the penetration rate increases to 10\%, our model still has better performance than the baseline models. Although the macro-micro model achieves 14\% higher performance than ours in average value while the penetration rate is 15\%, our model has a more stable performance. \textbf{\textit{In terms of speed accuracy,}} the performance is similar as the one of time headway. \textbf{\textit{In terms of fuel consumption and CO$_2$ emissions,}} we only list the detailed value of fuel consumption since CO$_2$ has a linear relationship with fuel consumption. Our model with MFC reconstruction method achieves a higher accuracy than the baseline models under all of the penetration rates. Besides the average value, it also provides a more stable performance than others. 
Moreover, we analyze the reason for the higher error of the polynomial reconstruction method. The fuel consumption of the reconstructed trajectory is always lower than the one of the ground truth trajectory, which implies that the trajectory may be over-smooth to lose some randomness existing in the ground truth trajectory. The randomness may largely affect the performance in terms of energy consumption. In summary, our model has a better performance, especially when the CV penetration rate is low.

\begin{figure}[!h]
    \centering
    \includegraphics[width=0.9\textwidth]{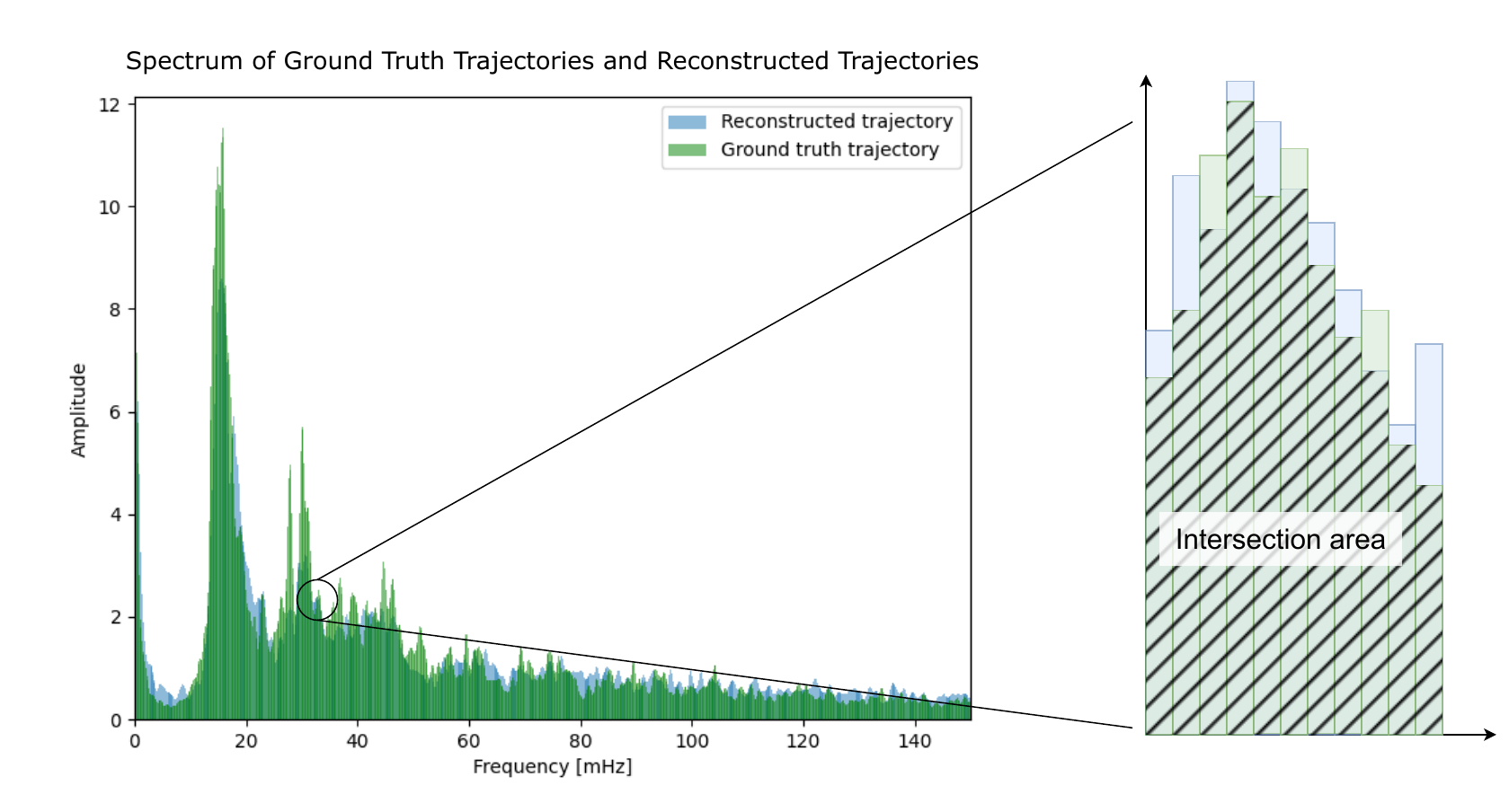}
    \caption{Evaluation metric for reproducing traffic oscillation.}
    \label{fig:oscillation_score}
\end{figure}

\subsection{Traffic Oscillation Reproduction}
\begin{figure}[htb!]
    \centering
    \includegraphics[width=0.85\textwidth]{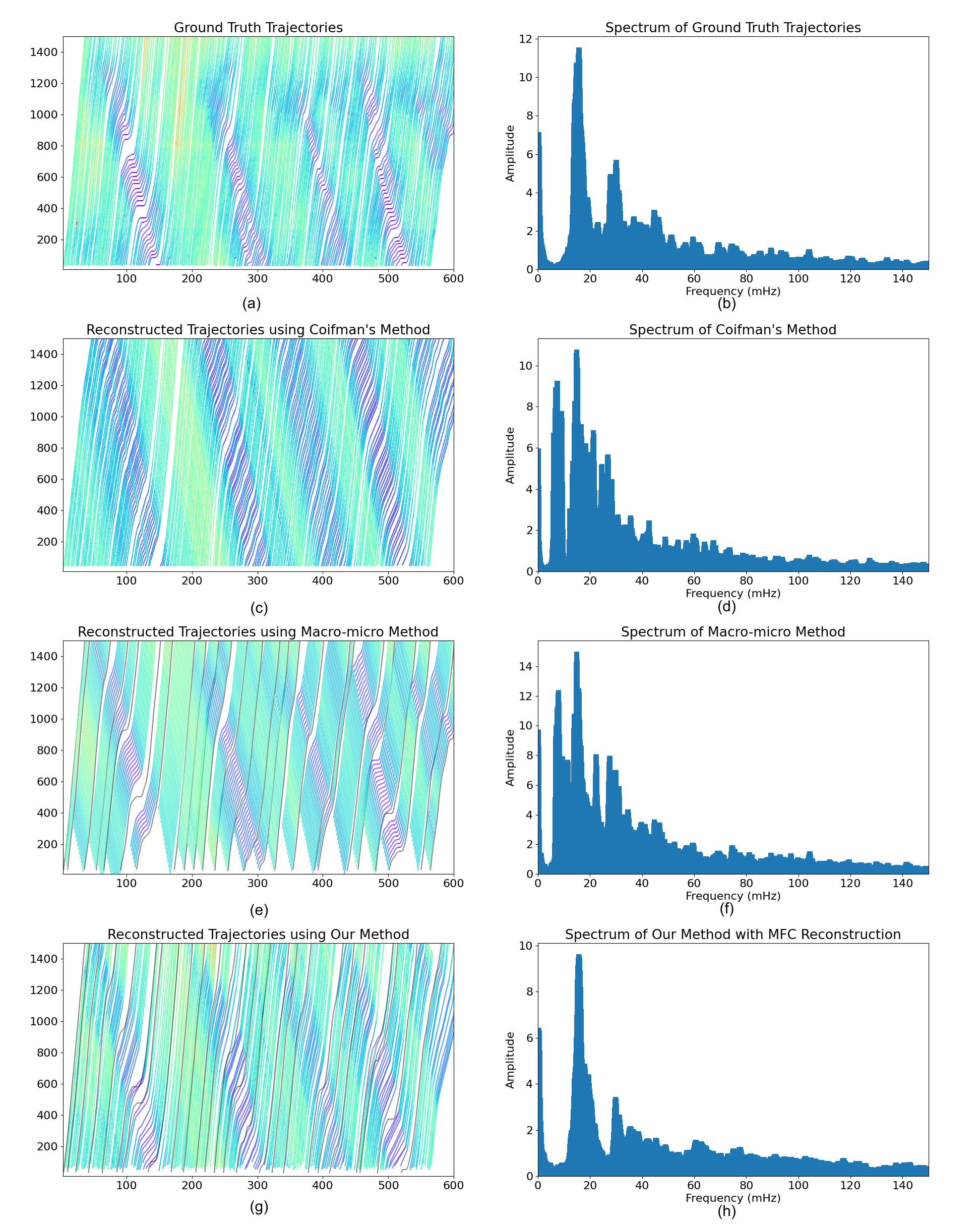}
    \caption{The Fourier Spectrum of ground truth data and reconstructed data. (a) Ground truth trajectories. (b) Fourier spectrum of ground truth. (c) Reconstructed trajectories by Coifman's method. (d) Fourier spectrum of reconstructed trajectories of Coifman's method. (e) Reconstructed trajectories by macro-micro method, where the CV penetration rate is \textbf{10\%} and the black line is the trajectory of CV. (f) Fourier spectrum of reconstructed trajectories of macro-micro method. (g) Reconstructed trajectories by our method with MFC reconstruction, where the CV penetration rate is \textbf{10\%} and the black line is the trajectory of CV. (h) Fourier spectrum of reconstructed trajectories of our method. }
    \label{fig:spectrum}
\end{figure}

In this part, we evaluate the capability of the proposed methodology in reproducing observed traffic oscillations. To characterize the traffic oscillations, we use the frequency and amplitude of the velocity sequence since the velocity can well reflect empirical observations according to~\citep{traffic_oscillation_freq_amp}. To this end, we apply the efficient Fast Fourier Transform (FFT) on the velocity sequence to get the frequency domain information. We use one of the subdatasets to illustrate the results. As shown in Fig.~\ref{fig:spectrum}, Fig.~\ref{fig:spectrum}.(b) is the spectrum of Fig.~\ref{fig:spectrum}.(a), which is expressed using frequency and amplitude. Specifically, except for the dominant peak amplitude, the existence of other peak amplitudes implies multiple traffic oscillations~\citep{traffic_oscillation_freq_amp}. We apply FFT to the reconstructed trajectory of Coifman's method, macro-micro method, and ours with MFC reconstruction. The reconstructed trajectory and the corresponding Fourier spectrum are shown in Fig.~\ref{fig:spectrum}.(c)-(h), respectively. Comparing Fig.~\ref{fig:spectrum}.(b), (d), (f), and (h), our model has the most similar peak frequency and amplitude as the one of ground truth, which indicates that our reconstruction method can successfully reproduce the traffic oscillations.

Moreover, considering that the connected vehicles are selected randomly in the above evaluation experiments, the spectrum may be affected by the selection of connected vehicles. Thus, we introduce a new metric to quantify the capability of traffic oscillation reproduction. Specifically, we calculate the intersection area between the spectrum of ground trajectories and reconstructed trajectories. The method of calculating the intersection area is shown in Fig.~\ref{fig:oscillation_score}. Then, we use the ratio of the intersection area on the whole area of the reconstructed spectrum to measure the quality of the reconstructed trajectories in terms of reproducing traffic oscillation. The ratio of Coifman's method is \textbf{77.20\%}. The mean and standard deviation value of the ratio achieved by macro-micro model is \textbf{63.90\%} and \textbf{1.01\%}, respectively. For our model with MFC reconstruction method, we achieve \textbf{88.61\%} and \textbf{1.41\%}, in terms of mean and standard deviation value of the ratio. Thus, compared with other baseline models, our model is the most capable one to reproduce the traffic oscillations.

\section{Discussion} \label{sec:discussion}
As the above experimental results illustrate that the accuracy of the reconstructed trajectory is affected by the characteristics of the CVs' trajectory. To well measure the contribution of different CVs, we propose a new metric to evaluate the contribution of the CV to trajectory reconstruction, which is spatial-temporal area coverage. Further analysis is conducted to measure the amount of information required to calibrate different CVs. By evaluating these two proposed metrics, we can identify the most critical and informative CV.

\subsection{Spatial-Temporal Area Reconstruction}
\begin{figure}[htbp]
    \centering
    \subfigure[An example of the reconstructed ST area. The unit of ST area is feet$\cdot$s.]{
    \includegraphics[width=0.35\textwidth]{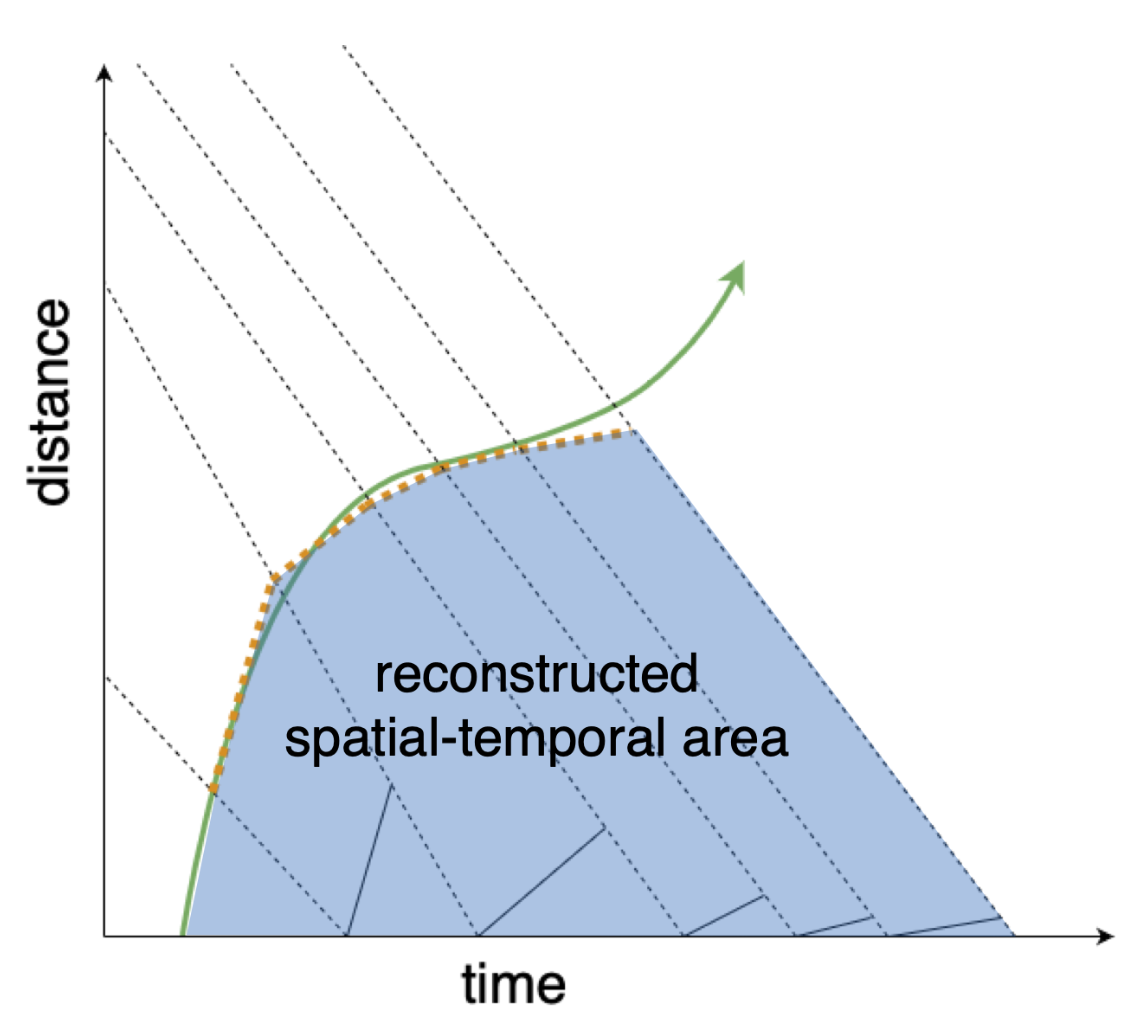}
    \label{fig:st_area_eg}
    }
    \subfigure[Relationship between the average speed of the CV and its reconstructed spatial-temporal area. The unit of y-axis is feet$\cdot$s and x-axis is feet/s.]{
    \includegraphics[width=0.5\textwidth]{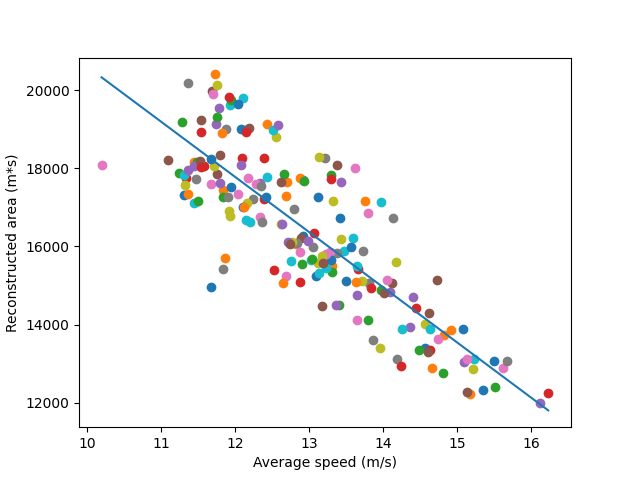}
    \label{fig:st_area_relation}
    }
    \caption{Analysis of the reconstructed spatial-temporal area of CVs.}
    \label{fig:st_area}
\end{figure}

Considering that the trajectory of the vehicles that are farther from the CV can not be fully recovered due to the limited number of calibrated shockwave speed values, we propose to reconstruct a spatial-temporal (ST) area instead of a set of complete trajectories. Given a set of calibrated shockwave speed values for a CV, the covered ST area of this CV is defined as the area enclosed by the reconstructed trajectory of the CV and the last shockwave line that can contribute to recovering the trajectory of other NCVs, as shown in Fig.~\ref{fig:st_area_eg}. Note that the trajectory of an NCV is considered as reconstruction completion when the reconstructed segment is more than $80\%$ of the whole journey. The larger of this area represents that knowing the trajectory of this CV can contribute more to trajectory reconstruction.

Given the above definition, we explore the relationship between the average speed of the CV and its covered ST area. We calibrate the time-varying shockwave speed for all vehicles. The results are shown in Fig.~\ref{fig:st_area_relation}, which indicates that the CV with a smaller average speed can cover more ST area and contribute more to reconstructing the trajectory of its following vehicles.

\subsection{Information Amount Requirement for Calibration}
\begin{figure}[htbp]
    \centering
    \includegraphics[width=0.55\textwidth]{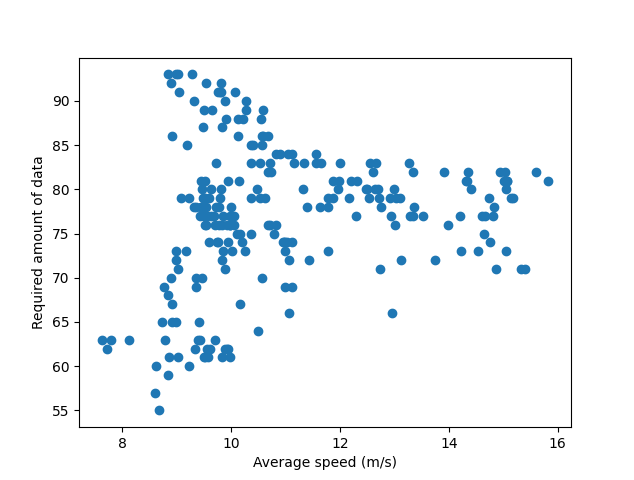}
    \caption{Relationship between the average speed of CVs and the required amount of data for shockwave calibration.}
    \label{fig:info_req}
\end{figure}

As introduced in the previous sections and experiments, the collected data, including arrival time and speed of following vehicles, is required to calibrate the time-vary shockwave speed values of the leading CV. However, for different CVs, they need different amount of data for calibration. We define the number of following vehicles' speed that are used to calibrate the shockwave speed of the CV as the information amount required for calibrating the time-varying shockwave speed. As shown in Fig.~\ref{fig:info_req}, there is no significant relationship between the average speed of the CV and its required amount of information. Thus, the selection of a CV has little impact on the efficiency and effectiveness in terms of the data requirement.

\section{Conclusion and Future Work} \label{sec:conclusion}
In conclusion, our proposed framework presents a comprehensive solution to the challenges in reconstructing vehicle trajectories using a combination of fixed and connected vehicle sensor data. The integration of fixed sensor information and connected vehicle data enables a more accurate and robust trajectory reconstruction process. The introduced calibration algorithm for time-varying shockwave speed further enhances precision, addressing the limitations of existing methods. Moreover, the vehicle dynamics are considered while reconstructing the trajectory for better energy estimation. Through real-world evaluations, our method consistently outperforms baseline models, demonstrating its effectiveness across various traffic conditions, especially in scenarios with a low penetration rate of CVs. The concept of a reconstructed ST area and information amount requirement provide valuable insights into the contribution and effectiveness of CVs with different average speeds, enriching our understanding of the reconstruction process. This work not only advances the field of trajectory reconstruction but also lays the foundation for more accurate modeling and optimization in traffic flow management and related domains. 

In future work, we plan to apply our method to traffic scenarios with signals and intersections to measure the queue length of intersections. Moreover, we will explore and build the relationship between the time-varying shockwave speed and traffic data collected by a loop detector using deep learning methods. As such, we can reconstruct the trajectories without knowing any CV trajectory, which can significantly enlarge the application scenarios and mitigate the reliance on the existence of CVs.


\printcredits

\bibliographystyle{cas-model2-names}

\bibliography{cas-refs.bib}

\bio{}
\endbio

\end{document}